\documentclass[12pt]{article}
\usepackage{latexsym,graphicx}
\usepackage{amsmath, amssymb, bm,eucal}
\begin{document}
\begin{flushright}
USTC-ICTS-13-06\\
\end{flushright}
\vspace{20mm}

\begin{center}

{\Large \bf Sakai-Sugimoto model in D0-D4 background}

\vspace{10mm}

{\large Chao Wu$^a$ \footnote{Email:wuchao86@mail.ustc.edu.cn}, Zhiguang
Xiao$^{a,b}$ \footnote{Email:xiaozg@ustc.edu.cn}, and Da Zhou$^a$
\footnote{Email:zhouda@mail.ustc.edu.cn}
}

\vspace{4mm}

{\em
 $^a$ Interdisciplinary Center for Theoretical Study\\
 University of Science and Technology of China, Hefei, Anhui
 230026, China\\

\vspace{4mm}

 $^b$State Key Laboratory of Theoretical Physics,
\\
 Institute of Theoretical Physics, Chinese Academy of Sciences 
\\}

\end{center}

\vspace{10mm}
%
%
%
%

\abstract{We add smeared D0 charges to the D4 background and
discuss Sakai-Sugimoto model under this background. The corresponding
gauge theory develops a  glue condensate $\langle tr (F_{\mu\nu}\tilde
F^{\mu\nu})\rangle $. The D8 branes go less deep than in the
original S-S model and massless Goldstones are still found in the
spectrum. The effects of the condensate on the meson
spectra, pion decay constant, and couplings of the vector mesons and
Goldstones are then investigated.  }

\section{Introduction}

Confinement as a nonperturbative phenomenon of QCD, attracts
lots of attention from theoretical physicists. There are many
mechanisms proposed to be the  possible cause of the confinement. See
\cite{Simonov:1997js} for a review. Among these mechanisms, some 
classical or semi-classical gauge field configurations could play an
important role, such as some topologically nontrivial
solutions---monopoles, instantons, and so on.  There could also be
solutions with constant field strength for the classical
equation of motion.  Selfdual field strength is studied in
\cite{Leutwyler:1980ev,Minkowski:1981ma,Flory:1983dx,vanBaal:1984ar},
and was proposed to be a mechanism for the confinement
\cite{Efimov:1998hi}. So there could be states with nonzero $tr
(F_{\mu\nu}\tilde F^{\mu\nu})$ background where $F_{\mu\nu}$ is the field
strength and $\tilde F^{\mu\nu}$ its dual, and they may play a role in
the confinement.

Such kind of states may also have some possibilities of being produced in
the heavy ion collisions.  There was some proposals that $P/CP$ odd bubble 
may be created during the collisions
\cite{Kharzeev:1998kz,Buckley:1999mv,Kharzeev:2004ey}. Some metastable
state with non-zero QCD vacuum $\theta$ angle or $tr (F_{\mu\nu}\tilde F^{\mu\nu})$ could
be produced in some space-time region in the hot and dense condition
when deconfinement happens. Then as the rapid expansion of the bubble,
it cools down and the metastable state freezes inside the
bubble\cite{Buckley:1999mv}. Then a $P$ or $CP$ odd bubble  may form.
It will soon decay into the true vacuum.

As nonperturbative phenomena in QCD, the effects of the states with
nonzero $tr (F_{\mu\nu}\tilde F^{\mu\nu})$ must be studied using
nonpertubative methods. String/gauge duality
provides a way to study this kind of phenomena. To add $\langle tr
(F_{\mu\nu}\tilde F^{\mu\nu})\rangle $ condensate in N=4 SUSY YM
corresponds to adding smeared D(-1) charges into D3 brane
configuration. Supersymmetric(SUSY) D(-1)-D3 background was  studied in
\cite{Liu:1999fc,Kehagias:1999iy} and was proposed to correspond to
gauge field theory with a selfdual background field strength
\cite{Liu:1999fc}. Non-SUSY D(-1)-D3 was studied in
\cite{Kehagias:1999iy,Ghoroku:2005tf} and corresponds to adding a
temperature to the corresponding gauge theory. By introducing D7 probe
branes into the background geometry, under the proposal of Karch and
Katz\cite{Karch:2002sh}, flavors can also be introduced into these
backgrounds, and then quark condensates, meson spectra could be
studied\cite{Ghoroku:2004sp,Ghoroku:2005tf,Brevik:2005fs,Ghoroku:2006af,Erdmenger:2007vj,Erdmenger:2011sz}.
Also by introducing baryonic D5 branes, studies on baryon properties
in the glue condensates can be carried out
\cite{Ghoroku:2008tg,Ghoroku:2008na,Sin:2009yu,Gwak:2012ht,Sin:2009dk}.

Another holographic construction of the QCD like theory is to use D4
background initiated by Witten \cite{Witten:1998zw}. By compactifying
the D4 brane on a circle, four dimensional Yang-Mills theory can be
obtained from the five dimensional Yang-Mills theory, and by imposing
the antiperiodic boundary condition on the fermions, supersymmetry is
broken. Flavors can be added into the Yang-Mills by introducing flavor
D6 \cite{Kruczenski:2003uq} or D8 branes \cite{Sakai:2004cn}. In
particular, Sakai and Sugimoto(S-S) \cite{Sakai:2004cn} proposed a model with D8-$\overline {\rm
D8}$ probe branes, where the spontaneous breaking of chiral symmetry
is geometrically realized as the joining of $N_f$ D8 branes and $N_f$ anti-D8 branes
into $N_f$ D8 branes at the tip. Massless Goldstones with the right
quantum numbers can be found in the spectrum. Meson spectra and
interactions then can be studied along these lines \cite{Sakai:2005yt}. Baryons
can also be easily realized as instantons in this model such that the nucleon
interactions can also be
modelled\cite{Hata:2007mb,Hashimoto:2008zw,Hashimoto:2009ys,Kaplunovsky:2010eh}.
As in the D(-1)-D3 background, adding condensate $\langle tr
(F_{\mu\nu}\tilde F^{\mu\nu})\rangle$ in the gauge theory corresponds
to adding smeared $D0$ charges into
the $D4$ background. The gauge theory in this background is studied in
\cite{Barbon:1999zp,Suzuki:2000sv}. Putting Sakai-Sugimoto model (S-S
model) into this background  allows us to study the hadron
phenomena in the nonzero $\langle tr (F_{\mu\nu}\tilde F^{\mu\nu})\rangle$ background.  In the
present paper, as a first step, we study the meson spectra and the
interactions of the lowest-lying
vector mesons and Goldstones in this
background. To keep the $\langle tr (F_{\mu\nu}\tilde F^{\mu\nu})\rangle$ dependence in the
large $N_c$, we require it to be of ${\mathcal O}(N_c)$ as in \cite{Liu:1999fc}, $\tilde
\kappa\sim\langle
tr (F_{\mu\nu}\tilde F^{\mu\nu})\rangle/ N_c $. There are 
still  massless Goldstone modes indicating the massless nature of
the flavor quarks. We analyse the lowest-lying scalar and vector meson spectra in this
model and the three point couplings for the lowest-lying vector mesons and Goldstones,
and find out that $\tilde \kappa$ really enters the formulae for these quantities. The detailed results are presented in section
\ref{sect:Scalar}, \ref{sect:Vector} and \ref{sect:Nonabelian}.

This paper is organized as follows: In section \ref{sect:D0D4}, we
review the D0-D4 background and its relation to the gauge field
theory. In section \ref{sect:D0D4D8} we put D8 probe branes into this
background and study the stability of the configuration. In section
\ref{sect:Scalar} and \ref{sect:Vector} we study the scalar and vector
meson spectra with one flavor, respectively. In section \ref{sect:Nonabelian}, we
extend our discussion to the multiflavor case, and  the interactions
of vector
mesons and Goldstones are studied. Section \ref{sect:conclude} is the
conclusion and discussion.

\section{The D0-D4 background\label{sect:D0D4}}
Some of the results in this section are already presented in \cite{Barbon:1999zp}. The solution of  D4 branes with smeared D0 charges in Type IIA
supergravity in Einstein frame is \cite{Barbon:1999zp,Suzuki:2000sv}
\begin{eqnarray}
ds^2&=& H_4^{-\frac3 8}\left(- H_0^{-\frac 7 8} f(U) d\tau^2+H_0^{\frac 1
8}\Big((dx^0)^2+(dx^1)^2+\cdots+(dx^3)^2\Big)\right)
\nonumber\\&&
+ H_4^{\frac5 8}H_0^{\frac 1 8}\left(\frac{dU^2}{f(U)}+U^2
d\Omega_4^2\right)\,,
\\
e^{-(\Phi-\Phi_0)}&=&(H_4/H_0^3)^{\frac 1 4}\,,
\\
f_2&=&\frac A {U^4} \frac 1{H_0^2}dU\wedge d\tau\,,
\\
f_4&=&B\epsilon_4\,,
\end{eqnarray}
where 
\begin{eqnarray}
A=\frac{(2\pi \ell_s)^7 g_s N_0}{\omega_4 V_4}\,&,& \quad B=\frac{(2\pi\ell_s)^3N_c g_s}{\omega_4}\,,
\\ 
H_4=1+\frac {U_{Q4}^3}{U^3}\,&,& \quad H_0=1+\frac{U_{Q0}^3}{U^3}\,,
\quad f(U)=1-\frac{U_{KK}^3}{U^3}\,.
\end{eqnarray}
$d\Omega_4$, $\epsilon_4$, and $\omega_4=8\pi^2/3$ are the line
element, the volume form and the volume of a unit $S^4$. $U_{KK}$ is the
coordinate radius of the horizon,
and $V_4$ the volume of D4-brane. $N_0$ and $N_c$ are the numbers
of D0  and D4 branes, respectively. D0 branes are smeared in the
$x^0,\dots, x^3$ directions.

In string frame the metric reads
\begin{eqnarray}
ds^2&=& H_4^{-\frac1 2}\left(- H_0^{-\frac 1 2} f(U) d\tau^2+H_0^{\frac 1
2}dx^2\right)+
H_4^{\frac1 2}H_0^{\frac 1 2}\left(\frac{dU^2}{f(U)}+U^2 d\Omega_4^2\right)
\end{eqnarray}
where $dx^2= (dx^0)^2+(dx^1)^2+\cdots+(dx^3)^2$ is used. The EOM requires:
\begin{eqnarray}
 A^2=9 U_{Q0}^3(U_{Q0}^3+U_{KK}^3) \,, \quad B^2=9U_{Q4}^3(U_{Q4}^3+U_{KK}^3)
\end{eqnarray}
which can be solved
\begin{eqnarray}
\label{eq:UQ0} 
U_{Q0}^3=\frac 1 2 (-U_{KK}^3+ \sqrt{U_{KK}^6+\frac 4 9 A^2})\,,
\\
U_{Q4}^3=\frac 1 2 (-U_{KK}^3 + \sqrt{U_{KK}^6+\frac 4 9 B^2})\,.
\end{eqnarray}
We have required $U_{KK}$ to be the horizon and no bare singularity,
and then 
$U^3_{Q0}>0, U^3_{Q4}>0$ are chosen. To use this solution in the
Sakai-Sugimoto model, we make a double wick rotation in
$\tau$ and $x^0$ directions and the metric becomes:
\begin{eqnarray}
ds^2&=& H_4^{-\frac1 2}\left(H_0^{-\frac 1 2} f(U) d\tau^2+H_0^{\frac 1
2}dx^2\right)+
H_4^{\frac1 2}H_0^{\frac 1 2}\left(\frac{dU^2}{f(U)}+U^2 d\Omega_4^2\right)
\nonumber \\
\label{eq:metic-Min}
\end{eqnarray}
where $dx^2= -(dx^0)^2+(dx^1)^2+\cdots+(dx^3)^2$ now.  In fact, the
metric is a bubble geometry and the space-time 
ends at $U=U_{KK}$.

In order not to have the conical singularity, the period of $\tau$
should be 
\begin{eqnarray}
\beta=\frac {4\pi}3 U_{KK}H_0^{1/2}(U_{KK})H_4^{1/2}(U_{KK})\,.
\end{eqnarray}
 We can then define a Kaluza-Klein mass scale $M_{KK}= 2\pi/\beta$, which
indicates the UV cut-off of the gauge theory. The D4 brane tension can
be related to the five dimensional Yang-Mills
coupling constant:
\begin{eqnarray}
\frac 1 {g_5^2}=\frac { (2\pi\alpha')^2}{(2\pi)^4\ell_s^5 g_s}=\frac 
{1}{(2\pi)^2\ell_s g_s}\,.
\end{eqnarray}
Then, by dimensional reduction to four dimensions, the four-dimensional
Yang-Mills coupling constant can be expressed as
\begin{eqnarray}
\frac 1{g_{YM}^2}=\frac{\beta}{g_5^2}=\frac \beta{4\pi^2 g_s
\ell_s}\,.
\end{eqnarray}
In another way, the string coupling constant can be expressed using
gauge theory parameters
\begin{equation}
 g_s=\frac {g_{YM}^2}{2\pi M_{KK} \ell_s}=\frac {\lambda}{2\pi
M_{KK}N_c \ell_s}\,,
\end{equation}
where $\lambda=g_{YM}^2 N_c$ is the 't Hooft coupling. Substituting this into (\ref{eq:UQ0}), we have
\begin{eqnarray}
H_0(U_{KK})= \frac 1 2 (1+ (1+ C \beta^2)^{1/2})\,,\quad
C\equiv(2\pi\ell_s^2)^6 \lambda^2\tilde \kappa^2/U_{KK}^6\,.
\label{eq:H0}
\end{eqnarray}
In order to keep the backreaction of D0 brane, we require $N_0$ to be
of order $N_c$ as in  \cite{Liu:1999fc} and define $\tilde \kappa=N_0/(N_c V_4)$. It is easy to
see that  $H_0(U)\ge 1$.

Going to the near horizon limit by taking $U/\alpha'$ and
$U_{KK}/\alpha'$ finite, we have
\begin{eqnarray}
U_{Q4}^3&\to& \pi \alpha'^{3/2} g_sN_c = \frac {\beta g_{YM}^2 N_c
\ell_s^2}{4\pi}\equiv R^3\,,
\label{eq:R3}
\\
H_4(U_{KK}) & \to & \frac{R^3}{U_{KK}^3}\,,
\\
\beta &\to& \frac {4\pi} 3 U_{KK}^{-1/2}R^{3/2} H_0^{1/2}(U_{KK})\,,
\label{eq:beta}
\\
M_{KK}&\to&
\frac 3 2 U_{KK}^{1/2}R^{-3/2}H_0^{-1/2}(U_{KK})\,.
\end{eqnarray}
The metric in string frame  then becomes
\begin{eqnarray}
ds^2&=&\left(\frac U R\right )^{3/2} \left(H_0^{1/2}(U) \eta_{\mu\nu}dx^\mu
dx^\nu+ H_0^{-1/2}(U) f(U)d\tau^2\right )
\nonumber \\&&+H_0^{1/2}\left(\frac R
U\right)^{3/2}\left(\frac 1{f(U)} dU^2+U^2 d\Omega_4^2\right)\,,
\end{eqnarray}
and the dilaton
\begin {eqnarray}
e^{\Phi}&=&g_s\left(\frac U R\right)^{3/4}H_0^{3/4}\,.
\end{eqnarray}
From (\ref{eq:R3}) and (\ref{eq:beta}) we have 
\begin{equation}
\beta^{1/2}=\frac 2 3 \pi^{1/2} U_{KK}^{-1/2}\lambda^{1/2}\ell_s
H_0^{1/2}(U_{KK})\,,
\end{equation}
or
\begin{equation}
\beta =\frac {4\pi\lambda \ell_s^2} {9  U_{KK}} H_0(U_{KK})\,,\quad
M_{KK}=\frac 9 2 \frac {U_{KK}}{\lambda \ell_s^2 H_0(U_{KK})}\,.
\label{eq:beta-H0}
\end{equation}
Since $H_0(U_{KK})\ge 1$, $U_{KK}\ge 2\lambda\ell_s^2M_{KK}/9$.

\begin{figure}[!t]
\centerline{\includegraphics[width=7cm]{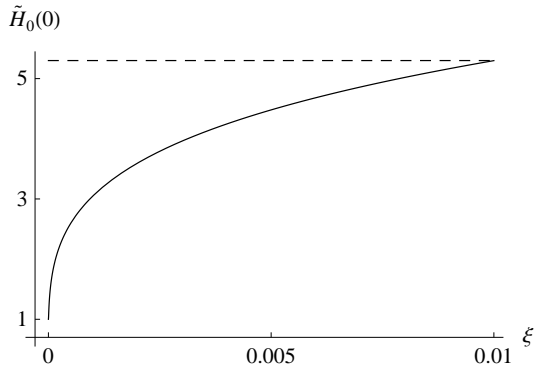}}
\caption{The relation between $\tilde{H}_0(0)$ and parameter $\xi$}
\label{fg:H0-xi}
\end{figure}

From this equation, $\beta$ can be solved and comparing with
(\ref{eq:beta-H0})
we have 
\begin{eqnarray}
\beta&=&\frac { 4\pi \lambda \ell_s^2}{9 U_{KK}} \frac 1{1-\frac{(2\pi
\ell_s^2)^8}{81 U_{KK}^8}\lambda^4\tilde \kappa^2}
\,,\quad H_0(U_{KK})=\frac 1{1-\frac{(2\pi
\ell_s^2)^8}{81 U_{KK}^8}\lambda^4\tilde \kappa^2}\,.
\label{eq:beta-M}
\end{eqnarray}
If we define 
$ D=\frac 2 9 \pi \lambda \ell_s ^2/U_{KK}$ and use the definition of
$C$ in (\ref {eq:H0}), $\beta$ then can be
expressed as $\beta=2D/(1-CD^2)$ and $H_0(U_{KK})=1/(1-CD^2)$. 
Since $H_0>0$ and $CD^2\leq 1$, this gives a constraint for $\tilde\kappa$, 
\begin{equation}
|\tilde \kappa|
\leq \frac{9U_{KK}^4}{ (2\pi \ell_s)^4\lambda^2}=\frac{\lambda^2 M_{KK}^4
H_0^4(U_{KK})}{9^3\pi^4}.\label{eq:bound}
\end{equation}
If we fix $\beta$, $\lambda$, from (\ref{eq:beta-H0}), $U_{KK}$
goes the same as $H_0(U_{KK})$. And together with (\ref{eq:beta-M}),
$H_0(U_{KK})$ and $\tilde \kappa$ can be
related 
\begin{eqnarray}
H^8_0(U_{KK})-H_0^7(U_{KK})=\frac{9^6\pi^8\tilde \kappa^2}{\lambda^4
M^8_{KK}}=9^6\pi^8\xi^2\,.
\label{eq:H0-kappa}
\end{eqnarray}
For future convenience, we have defined a dimensionless quantity
$\xi$
\begin{eqnarray}
\xi \equiv \frac{|\tilde{\kappa}|}{\lambda^2 M_{KK}^4}\,.
\label{eq:xi}
\end{eqnarray}
Since we fix $\lambda$ and $M_{KK}$, changing $\tilde \kappa$ is
equivalent to changing $\xi$. 
The left-hand side of (\ref{eq:H0-kappa}) is a monotonic function  increasing from zero for
$H_0(U_{KK})\geq 1$. So for each $\tilde \kappa$, there
is only one solution of $H_0(U_{KK})$, going up as $\tilde \kappa$
increases (see Figure \ref{fg:H0-xi}), and
$U_{KK}$ is similar. Since we are interested in the region with
$\lambda\gg 1$, if we choose $\lambda \sim 10$ and
$|\tilde\kappa|<M_{KK}^4$, $\xi$ should be within $0<\xi<0.01$. And the
corresponding $H_0(U_{KK})$ falls in $1<H_0(U_{KK})<5.3$. So in future
numerical analysis we constraint ourselves in this region.

This background actually introduces another free parameter $\tilde\kappa$ in the
Sakai-Sugimoto model. This string theory background is not dual to the vacuum state of the gauge
theory. The dual state may describe some excited
state with some constant homogeneous field strength background which
gives the expectation
value of $tr (F_{\mu\nu}\tilde F^{\mu\nu})$. On the supergravity side, $\tilde \kappa N_c$ is
the flux of $f_2$. Since $C_1$ is coupled to $tr (F_{\mu\nu}\tilde F^{\mu\nu})$
in the Euclidean Chern-Simons action
\begin{eqnarray}
S_{CS}= i \frac {\mu_4} 2 (2\pi\alpha')^2 \int d\tau C_\tau\wedge
tr(F\wedge F)
\end{eqnarray}
$\tilde\kappa$ characterizes the expectation value of the Euclidean $tr (F_{\mu\nu}\tilde F^{\mu\nu})$. Just
as in \cite{Liu:1999fc}, by rotating to the Euclidean space and naively using classical EOM of $C_1$, we
have real Euclidean condensate
\begin{equation}
\langle tr (F_{\mu\nu}\tilde F^{\mu\nu})\rangle =8\pi^2N_c \tilde \kappa
\end{equation}
  We suppose this is a stochastic average over the
background fields in all directions so that the $\langle F\rangle$ is
still zero and the four dimensional space-time translation invariance and proper Lorentz
invariance are preserved which is manifest in the string background
solution. Obviously the $P$ and $CP$ invariances are
violated. This is just  similar to the situation in 
 \cite{Liu:1999fc} by H. Liu {\it et al}. Selfdual constant homogeneous backgrounds in
the gauge theory are studied
in
\cite{Leutwyler:1980ev,Minkowski:1981ma,Flory:1983dx,vanBaal:1984ar}
and may be related to the confinement. However, the field strength
may not be selfdual in the present paper since the gravity background
is non-supersymmetric, and so we will not take it as 
 a necessary assumption. Whether the background
field strength is selfdual or not is beyond the scope of this  paper.
 Our interest is to 
put S-S model in
this background to study the $\tilde
\kappa$ dependence of the meson spectra and couplings.

Now we have some independent parameters on the gravity side: $R^3$,
$U_{Q0}^3 $, $ U_{KK}$ and  $g_s$, and $\ell_s$ will be cancelled out
in the final physical results. We also have some parameters on the
gauge theory side $N_c$, $M_{KK}$, $\lambda$ and $\tilde \kappa$.  We
have seen that $\tilde \kappa$ can be related to $H_0(U_{KK})$ and we
can use $H_0(U_{KK})$ to represent $\tilde \kappa$. The final results
on the gauge theory side can be expressed using $N_c$, $M_{KK}$,
$\lambda$ and $H_0(U_{KK})$. We collect the relations here:
\begin{eqnarray}
R^3=\frac {\lambda \ell_s^2}{2M_{KK}}\,,\quad g_s=\frac
{\lambda}{2\pi M_{KK}N_c\ell_s}\,,\quad U_{KK}=\frac 2 9 M_{KK}\lambda
\ell_s^2 H_0(U_{KK})\,. 
\label{eq:constants}
\end{eqnarray}
We fix the gauge theory parameters $M_{KK}$, $N_c$ and $\lambda$, and then
change $\tilde \kappa$. This corresponds to fixing the parameters on
the
gravity side: $R^3$, $g_s$, $H_0(U_{KK})/U_{KK}$, and changing
$H_0(U_{KK})$ or $U_{KK}$.

Similar to the discussion of the D4-soliton background
\cite{Kruczenski:2003uq} in S-S model, we
can discuss the reliability of the background.  First we require
the curvature near the horizon to be small compared to the string
scale $1/|({\bf R}\ell_s^2)|\gg1$. The curvature at $U_{KK}$ is 
\begin{eqnarray}
{\bf R}(U_{KK})&\sim& \frac 9{R^{3/2}U_{KK}^{1/2}H_0^{1/2}(U_{KK})}
\left(2-\frac 3{H_0(U_{KK})}\right)
\end{eqnarray}
We have used $U_{KK}^3/R^3\sim \ell_s^6/\ell_s^2\to0$. Then using (\ref{eq:constants}), we have
\begin{eqnarray}
1\ll\left|\frac 1{{\bf
R}\ell_s^2}\right|\sim\left|\frac{R^{3/2}U_{KK}^{1/2}H_0^{1/2}(U_{KK})}{9\ell_s^2(2-3/H_0(U_{KK}))}\right|
\sim\left|\frac
{g_{YM}^2N_cH_0(U_{KK})}{27(2-3/H_0(U_{KK}))}\right|\,.
\end{eqnarray}
Since the factor $|H_0(U_{KK})/(27(2-3/H_0(U_{KK}))|\ge 1/27$ is bounded from below
for $H_0(U_{KK}) > 1$, $g^2_{YM}N_c\gg 1$ satisfies this inequality.
However, the denominator $(2-3/H_0(U_{KK}))$ could be zero for
$H_0$ near $3/2$. This may indicate that the gravity may not correspond to
strong coupling region. Nevertheless, by analysing the scalar
$R_{\mu\nu}R^{\mu\nu} \ell_s^4\ll 1$, we can conclude that near
$H_0=3/2$ the
corresponding gauge theory is really in the  strong 't Hooft coupling
region:
\begin{eqnarray}
1\ll 1/|R_{\mu\nu}R^{\mu\nu} \ell_s^4| \simeq \frac{\lambda^2
H_0^4(U_{KK})}{729(H_0^2-H_0+1)}\,.
\label{eq:R2}
\end{eqnarray}
However $H_0$ can not be arbitrarily large. We require the factor
$H_0^4/(729(H_0^2-H_0+1)$ to be of ${\cal O}(1)$,
which corresponds to 
 $H_0(U_{KK})\sim 30$. Notice that previously
$|\tilde \kappa| < M^4_{KK}$ and large $\lambda$ requires
$1\le H_0(U_{KK})<5.3$ which falls in this region.
So the smallness of the curvature corresponds to the large 't Hooft
coupling in the gauge theory in $1\le H_0<5.3$.

Next we require $e^\Phi\ll1$ to suppress the string loop effect. From 
$e^\Phi=g_s H_0^{3/4}H_4^{-1/4}$, we have 
\begin{eqnarray}
UH_0(U)\ll g_s^{-4/3}R
\end{eqnarray}
Since  $1\le H_0\sim {\cal O}(1)$, this means $U\ll g_s^{-4/3}R\equiv U_{\rm
crit}$. This introduces no new information to the D4 soliton
results. To repeat, the critical radius  
can be expressed as $U_{\rm
crit}\simeq(2\pi^{4/3}\ell_s^2N_c^{1/3}M_{KK})/g_{YM}^2$ and we
require $U_{\rm crit}\gg U_{KK}$. 
So we have 
\begin{equation}
g_{YM}^4\ll\frac 1{g_{YM}^2N_c} \ll1
\end{equation}
This just suggests that  the supergravity solution is a valid dual
description of the strong coupling region of the four
dimensional gauge theory in the 't Hooft limit.

\section{Sakai-Sugimoto model in D0-D4 background\label{sect:D0D4D8}}
Now we embed the D8 brane into the background with $U=U(\tau)$. The
metric then becomes
\begin{eqnarray}
ds^2&=&\left(\frac U R\right )^{3/2}  H_0(U)^{-1/2}\left( f(U)
+\left(\frac R
U\right)^{3}\frac {H_0(U)}{f(U)} {U'}^2\right )d\tau^2
\nonumber \\
&&+\left(\frac U R\right )^{3/2} H_0^{1/2}(U) \eta_{\mu\nu}dx^\mu
dx^\nu+H_0^{1/2}\left(\frac R
U\right)^{3/2}U^2 d\Omega_4^2
\end{eqnarray}
where $U'=dU/d\tau$.
Substitute this into D8-brane action, we have 
\begin{eqnarray}
S_{D8}\sim \frac 1{g_s}\int d^4x\,d\tau H_0(U) U^4 \left(f(U)+\frac
{H_0(U)}{f(U)}\left (\frac R U\right)^3 U'^2\right)^{1/2}\,,
\end{eqnarray}
from which the equation of motion can be obtained:
\begin{eqnarray}
\frac {d}{d\tau}\left(\frac{H_0(U)U^4f(U)}{\left[ f(U)+\frac
{H_0(U)}{f(U)}\left(\frac R U\right)^3U'^2\right]^{1/2}}\right)=0\,,
\end{eqnarray}
which is just the conservation of the energy.
With initial conditions $U(0)=U_0$ and $U'(0)=0$ at $\tau=0$,
$\tau(U)$ can be solved
\begin{equation}
\tau(U)=E(U_0)\int_{U_0}^{U} dU \frac
{H_0^{1/2}(U)\left(\frac R U\right)^{3/2}}{f(U)(H_0^{2}(U)U^8f(U)-E^2(U_0))^{1/2}}
\end{equation}
where $E(U_0)=H(U_0)U_0^4 f^{1/2}(U_0)$. 

\begin{figure}
\centerline{\includegraphics[width=7cm]{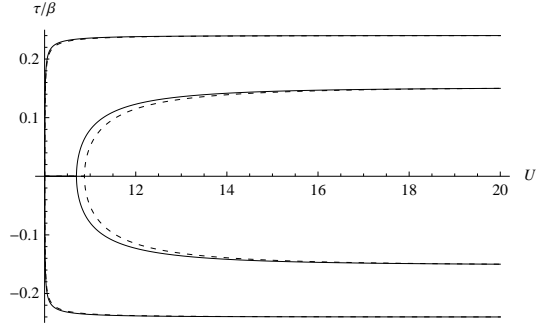}}
\caption{The profile of the D8 brane. The dashed lines denote the
profile with $\tilde \kappa \neq 0$ and the solid lines with zero
$\tilde \kappa$. With $\tilde \kappa\neq0$
, the D8 brane goes less deep.\label{fg:D8-profile}}
\end{figure}

The difference between the present background and the D4-soliton background is
the $H_0(U)$ factors in all the equations. If we set $H_0(U)\to 1$,
all the results degenerate to the original S-S model.
For the antipodal case the profile is the same as the original S-S
model, with $\tau(U) = \beta/4$. As the D8-$\overline {\rm D8}$ moves
away from the antipodes,
the profile goes less deep than in the original S-S model (Figure
\ref{fg:D8-profile}).
In this paper, as a first step, we constraint ourselves to the antipodal case to see the
 the effects of the condensate $\langle tr (F_{\mu\nu}\tilde F^{\mu\nu}) \rangle$.

 As in S-S model, we introduce the new coordinate $(r,\theta)$ or $(y,z)$:
\begin{eqnarray}
y=r \cos \theta\,&,& \quad z =r \sin \theta
\nonumber \\
U^3=U_{KK}^3+U_{KK} r^2\,&,& \theta =\frac {2\pi}{\beta}\tau=\frac 3 2
\frac{U_{KK}^{1/2}}{R^{3/2} H_0^{1/2}(U_{KK})}\tau
\end{eqnarray}
and  then the metric in the $(y,z)$ plane becomes
\begin{multline}
ds_{\tau,U}^2 = \frac 4 9 \frac{R^{3/2}}{U^{3/2}}
 \frac{H_0(U_{KK})}{H_0^{1/2}(U)}  \Bigg[ \Big(1-h(r)y^2\Big){dy^2} \\
+\quad \Big(1-h(r)z^2\Big){dz^2} -2yz h(r)dydz \Bigg]
\end{multline}
where
\begin{eqnarray}
h(r)=\frac{1}{r^2} \left[ 1-\frac{U_{KK}H_0(U)}{UH_0(U_{KK})}
\right]\,.
\end{eqnarray}
In the antipodal case, the D8 brane is put along $x^0, x^1, x^2, x^3$
and $z$ direction at $y=0$, wrapping the $S^4$. We can also study  the
fluctuations of D8 brane in the $y$ direction to examine the stability
of this configuration. Then $y$ is considered as a function of $x$ and $z$,
$y(x,z)$. The induced metric then reads 
\begin{eqnarray}
ds^2&=&ds^2_{\rm 5d}+
H_0^{1/2}(U)R^{3/2}{U}^{1/2} d\Omega_4^2\\
ds^2_{\rm 5d} &=& H_0^{1/2}(U)\left(\frac{U}{R}\right)^\frac{3}{2} \left[ \eta_{\mu\nu} +
\frac{4}{9} \left(\frac{R}{U}\right)^3 \frac{H_0(U_{KK})}{H_0(U)} \partial_\mu y\partial_\nu y
\right] dx^\mu dx^\nu \nonumber\\
&+& \frac{4}{9} \left(\frac{R}{U}\right)^\frac{3}{2} \frac{H_0(U_{KK})}{H_0^{1/2}(U)}
\left[ \frac{U_{KK}H_0(U)}{UH_0(U_{KK})} + \dot{y}^2 + h(z)\Big(y^2-2zy\dot{y}\Big) \right] dz^2 \nonumber\\
&+& \frac{8}{9} \left(\frac{R}{U}\right)^\frac{3}{2} \frac{H_0(U_{KK})}{H_0^{1/2}(U)}
\partial_\mu y \Big[ \dot{y}-zyh(z) \Big] dx^\mu dz + {\cal O}(y^4)
\end{eqnarray}
And the DBI action of D8-brane turns out to be
\begin{eqnarray}
S_{D8}&=&-\tilde T H_0^{3/2}(U_{KK})\int d^4 x dz \left [ \frac
{H_0^{3/2}(U_z)}{H_0^{3/2}(U_{KK})}U_z^2
\right.\nonumber \\&&\left.+\frac{H_0^{1/2}(U_z)}{H_0^{1/2}(U_{KK})}\left(\frac 2 9
\frac{R^3}{U_z}\eta^{\mu\nu}\partial_\mu y\partial_\nu y+\frac {U^3}{2U_{KK}}\dot
y^2+\frac 1 2\left(1+\frac 1{H_0(U_z)}\right)y^2\right)\right]
+{\cal O}(y^4)\,.
\nonumber \\
\end{eqnarray}
where we have defined $U_z=U_{KK}(1+z^2/U_{KK}^2)^{1/3}$, $ \tilde
T=\frac 2{3g_s} T_{8}\Omega_4U_{KK}^{1/2}R^{3/2}$, with $T_8=\left(
(2\pi)^8\ell_s^9 \right)^{-1}$ the tension of D8-brane.
Then the energy density of the fluctuations in the $y$ direction can be
read off
\begin{eqnarray}
{\cal E}&\simeq& \tilde T H_0^{ 3/2} (U_{KK})\int dz
\frac{H_0^{1/2}(U_z)}{H_0^{1/2}(U_{KK})}\left(\frac 2 9
\frac{R^3}{U_z}\sum_{i=0}^{3}(\partial_i y)^2+\frac {U^3}{2U_{KK}}\dot
y^2+\frac 1 2\left(1+\frac 1{H_0(U_z)}\right)y^2\right) 
\nonumber \\
&\geq& 0
\label{eq:energy}
\end{eqnarray}
So adding the D0 flux does not affect the stability of the D8
brane probe configuration with  respect to small fluctuations.

\section {Scalar meson spectrum\label{sect:Scalar}}
Using the results of the previous section, we are ready to discuss the
scalar spectrum for one  flavor case. The fluctuations of  $y$ can be
expanded in terms
of some orthogonal basis $\rho_n(z)$
\begin{eqnarray}
y(x^\mu,z) = \sum_{n=1}^{\infty} {\cal U}^{(n)} (x^\mu) \rho_n(z)\,.
\label{eq:y-expand}
\end{eqnarray}
 We now define the dimensionless $Z=z/U_{KK}$,
$K=1+Z^2=(U_z/U_{KK})^3$, $U^3_z=U_{KK}^3(1+Z^2)$ and $\tilde H_0(Z)=H_0(U_z)$.
The orthogonal condition for $\rho_m$ reads
\begin{eqnarray}
\frac 4 9 \tilde T R^3 \tilde H_0(0)\int dZ \tilde H_0^{1/2}(Z)
K^{-1/3}(Z)\rho_m\rho_n=\delta_{mn}\,,
\label{eq:rho-ortho}
\end{eqnarray}
and $\rho_m\ (m\ge1)$ are eigenfunctions of equation
\begin{multline}
K^{1/3}(Z)\left[ -\tilde H_0^{-1/2}(Z)\partial_Z
\left( \tilde H_0^{1/2}(Z)K(Z)\partial_Z\rho_n(Z) \right) \right.\\
\left. + \left( 1+\frac{1}{\tilde H_0(Z)} \right) \rho_n(Z) \right]
\ =\ \lambda_n \rho_n(Z)\,.
\label{eq:eigen-rho}
\end{multline}
Then the D8 action can be written as 
\begin{eqnarray}
S_{D8}=-\int d^4x \frac 1 2 \sum_{i=1}^{\infty}\partial_\mu{\cal
U}^{(n)}\partial^\mu{\cal U}^{(n)} +\frac 1 2 M_{KK}^2
\tilde H_0(0)\sum_n\lambda_n \left( {\cal U}^{(n)} \right)^2\,
\label{eq:D8-U}
\end{eqnarray}
from which we can read off the mass for scalar mesons 
\begin{eqnarray}
m^2_n=M_{KK}^2\tilde H_0(0)\lambda_n\,.
\label{eq:Mass-Scalar}
\end{eqnarray}
We see that the $\tilde \kappa$ dependence of the mass is through
$\tilde H_0(0)$ factor and is also hidden in $\lambda_n$ as a result of the eigenvalue
equation (\ref{eq:eigen-rho}).

Now, we proceed to solve the eigenvalue equation. Similar to the
method in Sakai and Sugimoto's original paper, from
(\ref{eq:eigen-rho}) we first  find out the asymptotic behavior of
$\rho_n(Z)$ as $Z$ goes to infinity:
\begin{eqnarray}
\rho_n \sim \frac{1}{Z^2}\,.
\end{eqnarray}
Then we can define
\begin{eqnarray}
Z \equiv e^\eta \ ,\qquad \tilde{\rho}_n(\eta) \equiv e^{2\eta} \rho_n(e^\eta)
\end{eqnarray}
such that $\tilde{\rho}_n$ is of $\mathcal{O}(Z^0)$. So the equation for $\tilde{\rho}_n$ reads
\begin{eqnarray}
\frac{d^2\tilde{\rho}_n}{d\eta^2} + G \frac{d\tilde{\rho}_n}{d\eta} + F \tilde{\rho}_n =0
\end{eqnarray}
where
\begin{eqnarray}
F &=& 6- \frac{3}{1+e^{-2\eta}} - \frac{3}{1+\tilde{H}_0(0)e^{-2\eta}}
+ \frac{\lambda_n e^{-2\eta/3}}{(1+e^{-2\eta})^{4/3}}
\equiv \sum^\infty_{k=0} F_k e^{-2k\eta/3}\,, \nonumber\\
G &=& -5+ \frac{1}{1+e^{-2\eta}} + \frac{1}{1+\tilde{H}_0(0)e^{-2\eta}}
\equiv \sum^\infty_{k=0} G_k e^{-2k\eta/3}\,.
\end{eqnarray}
The first few non-vanishing coefficients are listed below
\begin{eqnarray}
&& F_1=\lambda_n\ ,\ F_3=3+3\tilde{H}_0(0)\ ,\ F_4=-\frac{4}{3}\lambda_n\ ,\ \cdots \nonumber\\
&& G_0=-3\ ,\ G_3=-1-\tilde{H}_0(0)\ , \cdots
\end{eqnarray}
Next we expand $\tilde{\rho}_n$ as
\begin{eqnarray}
\tilde{\rho}_n\sim 1+ \sum^{\infty}_{k=1} \beta_k e^{-2k\eta/3}\,.
\end{eqnarray}
 And it is easy to verify that
\begin{eqnarray}
\beta_1=-\frac{9}{22}\lambda_n\ ,\ \beta_2=\frac{81}{1144}\lambda_n^2\ ,
\ \beta_3=-\frac{3}{10}-\frac{3}{10}\tilde{H}_0(0)-\frac{81}{11440}\lambda_n^3\ ,\ \cdots
\end{eqnarray}

We then solve the eigenvalue equation using ``shooting'' method with $\xi$ running
from 0 to 0.01. The same as in S-S's original paper \cite{Sakai:2004cn}, we choose the eigen-function
to be even or odd for $n\geq 1$
\begin{eqnarray}
\rm{Even: } & \partial_Z\rho_n(0)=0\,,\quad \rm{Odd: } &
\rho_n(0)=0\,.
\end{eqnarray}
\begin{figure}[!h]
\includegraphics[width=.5\textwidth]{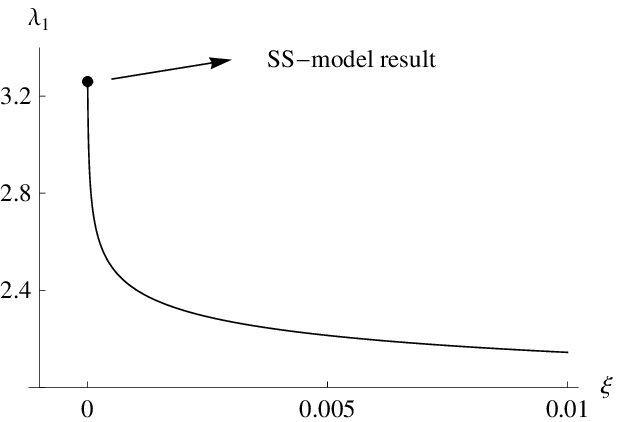}
\includegraphics[width=.5\textwidth]{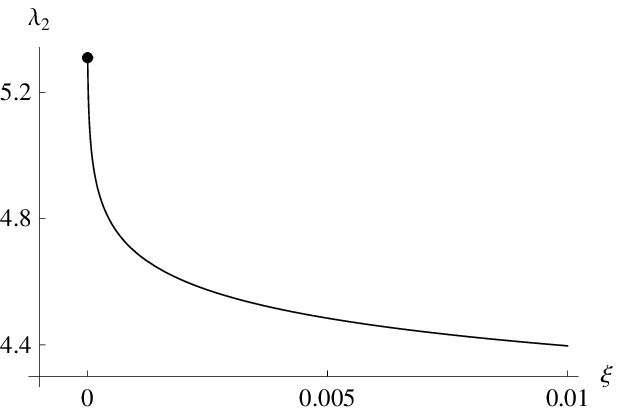}
\caption{The $\xi$-dependence of $\lambda_1$ and $\lambda_2$}
\label{fg:lambda-rho}
\end{figure}
\begin{figure}[!h]
\includegraphics[width=.5\textwidth]{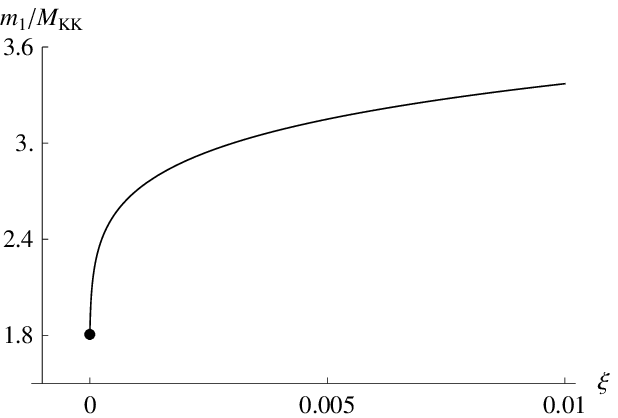}
\includegraphics[width=.5\textwidth]{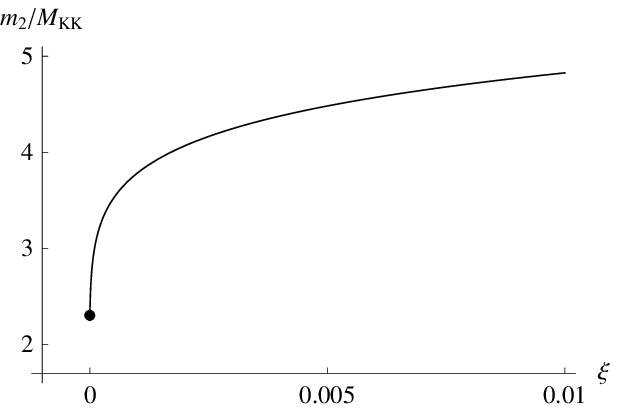}
\caption{The  $\xi$-dependence of $m_1$ and $m_2$}
\label{fg:mass-rho}
\end{figure}
As a result, the eigenfunctions are even for odd $n$, and odd for
even $n$.
The charge conjugate $C$ and
parity properties are the same as in S-S model. The the lightest scalar meson
has $CP=++$ and the next level $CP=--$. 
The $\xi$-dependence of the lowest two $\lambda_n$ and masses are
shown in Figure~\ref{fg:lambda-rho} and 
Figure~\ref{fg:mass-rho}, respectively. 

From these figures, we can see that even though the first two
eigenvalues go down as $\xi$ increases, the contributions from $\tilde
H_0(0)$ overcome the eigenvalue contributions and make the mass grow
with $\xi$.  This is different from the results using D(-1)-D3
background in \cite{Brevik:2005fs}. In their model, Liu-Tseytlin
\cite{Liu:1999fc} background is used, which is supersymmetric, and in
the corresponding gauge theory, the condensate is claimed to be
selfdual $\langle F_{\mu\nu}F^{\mu\nu}\rangle=\langle F_{\mu\nu}\tilde
F^{\mu\nu}\rangle\sim q$. Also the current quark mass is non-zero.  In
our model the background is not supersymmetric and it is highly
possible that the field strength is not selfdual. And since the
Goldstone is massless, the current quark mass is also zero. So the
difference is not surprising. In their model the meson mass is only
determined by the eigenvalue of the fluctuation which is going down
with increasing $q$. In our model, though the eigenvalues have the
same tendancy as theirs, the masses are also proportional to $\tilde
H_0(0)$ which is increasing and dominates in the contributions.

\section{Gauge field fluctuations and vector meson
spectra\label{sect:Vector}}
Now we consider the gauge field excitations on the D8 brane in this
background. As in S-S model, we are only interested in the SO(5)
singlets, $A_\mu$ ($\mu=0,1,2,3$) and $A_z$ which are independent of
the angular coordinates of the $S^4$.
We consider only one flavor in this section. The DBI action can be
cast into
\begin{eqnarray}
S_{D8}=-\tilde T(2\pi\alpha')^2\int d^4xdz\, H_0^{1/2}(U)\left[\frac 1 4 \frac
{R^3}U F_{\mu\nu}F^{\mu\nu}+\frac 9 8
\frac{U^{3}}{U_{KK}}F_{\mu z}F^{\mu z}\right]\,.
\end{eqnarray}
As in S-S model, we expand the gauge field $A_\mu (\mu=0,1,2,3)$ and
$A_z$ in terms of some
orthogonal basis,
\begin{eqnarray}
A_\mu(x,z)&=&\sum_{n=1}^{\infty} B_\mu^{(n)}(x)\psi_n(z)\,,
\nonumber\\
A_z(x,z)&=&\sum_{n=0}^{\infty} \varphi^{(n)}(x)\phi_n(z) \,,
\end{eqnarray}
and the orthogonal conditions are
\begin{eqnarray}
\tilde T(2\pi\alpha')^2 R^3\int dZ
\frac{\tilde H_0^{1/2}(Z)}{K^{1/3}(Z)}\psi_m\psi_n=\delta_{mn}\,,
\label{eq:ortho0}\\
\tilde T(2\pi\alpha')^2 R^3M_{KK}^2\tilde H_0(0)U_{KK}^2\int dZ
\tilde H_0^{1/2}(Z)K(Z)\phi_m\phi_n=\delta_{mn}\,.
\label{eq:ortho}
\end{eqnarray}
The eigenvalue equation for $\psi_m$ is
\begin{eqnarray}
-\tilde H_0^{-1/2}(Z) K^{1/3}(Z)
\partial_Z\left(\tilde
H_0^{1/2}(Z)K(Z)\partial_Z\psi_m\right)=\Lambda_m \psi_m\,,
\label{eq:eigeneq}
\end{eqnarray}
with $\Lambda_n$ the eigenvalue. The eigenfunction $\phi_n(z)$ can be chosen as 
\begin{eqnarray}
\phi_n=\frac{1}{M_n U_{KK}}\partial_Z \psi_n=\frac{1}{M_n}\dot
\psi_n(z)\,, \quad M_n=\Lambda_n^{1/2}M_{KK}\tilde H_0^{1/2}(0)\,,
\label{eq:phi}
\end{eqnarray}
for $n\neq 0$, and for $n= 0$
\begin{eqnarray}
\phi_0&=&\frac c {\tilde H_0^{1/2}(Z) K(Z)}\,,
\nonumber \\
c&=&\left(\tilde
T(2\pi\alpha')^2R^3M_{KK}^2\tilde H_0(0)U_{KK}^2\int
dZ\,\tilde H_0^{-1/2}(Z)K^{-1}(Z)\right)^{-1/2}
\nonumber\\
&=&\frac{9(3\pi)^{3/2}}{\sqrt{2\lambda N_c}\lambda M_{KK}^2 \ell_s^2
\tilde H_0^{3/2}(0)}\frac{1}{ \sqrt {{\cal F}_0}}
\label{eq:phi0}
\end{eqnarray}
with
\begin{eqnarray}
{\cal F}_0 \equiv  {\cal F} \left( \frac{\pi}{2} , \sqrt{1- \tilde{H}^{-1}_0(0)} \right) \,,\quad 
{\cal F}(\phi,k) &\equiv & \int_0^\phi
\frac{d\theta}{\sqrt{1-k^2\sin^2\theta}}
\nonumber
\label{eq:elliptic}
\end{eqnarray}
 the elliptic integrals of the first kind.
The DBI action of D8 brane can be recast into
\begin{eqnarray}
S_{D8} &=& -\int d^4x \left[\ \sum_{n=1}^\infty \left(
\frac{1}{4} F^{(n)}_{\mu\nu}F^{(n)\mu\nu}
+ \frac{1}{2} M_n^2 B_\mu^{(n)}B^{\mu(n)}
- M_n\partial_\mu\varphi^{(n)}B^{\mu(n)} \right)\right. \nonumber\\
&& \left. +\ \sum_{n=0}^\infty \frac{1}{2}
\partial_\mu\varphi^{(n)}\partial^\mu\varphi^{(n)}
\ \right]\,.
\end{eqnarray}
By replacing $B_\mu^{(n)}\to
B_\mu^{(n)}+M_n^{-1}\partial_\mu\varphi^{(n)}$ through
a gauge transformation, the action becomes
\begin{eqnarray}
S_{D8} = -\int d^4x \left[ \sum_{n=1}^\infty \left(
\frac{1}{4} F^{(n)}_{\mu\nu}F^{(n)\mu\nu}
+ \frac{1}{2} M_n^2 B_\mu^{(n)}B^{\mu(n)} \right)
+ \frac{1}{2} \partial_\mu\varphi^{(0)}\partial^\mu\varphi^{(0)}
\right]\,.
\label{eq:S-D8-gauge}
\end{eqnarray}

We see that the masses for the massive vector bosons are just $M_n$ and
there is a massless peudo-scalar $\varphi^{(0)}$ which is just the
Nambu-Goldstone boson. For $U(1)$ case here, this Goldstone boson is
just like the $\eta'$ in the real world. Due to the $U(1)_A$ anomaly,
its mass is related to the topological susceptibility of the pure
Yang-Mills theory. This has already been discussed in \cite{Barbon:1999zp} and
met some difficulties in obtaining the analytic results. So we will not go
deep in this direction.

We can now analyse the $\tilde \kappa$ dependence of the mass spectrum
of the vector mesons by performing the same procedure as in the previous section.
First we find out the asymptotic behavior when $Z$ approaches infinity
\begin{eqnarray}
\psi_n \sim \frac{1}{Z}
\end{eqnarray}
and define a new function
\begin{eqnarray}
\tilde{\psi}_n(\eta) = e^\eta \psi_n(e^\eta)
\end{eqnarray}
which satisfies the equation
\begin{eqnarray}
\frac{d^2\tilde{\psi}_n}{d\eta^2} + G'\frac{d\tilde{\psi}_n}{d\eta} + F'\tilde{\psi}_n =0
\end{eqnarray}
where
\begin{eqnarray}
F' &=& 2 -\frac{1}{1+e^{-2\eta}} -\frac{1}{1+\tilde{H}_0(0)e^{-2\eta}}
+ \frac{\Lambda_n e^{-2\eta/3}}{(1+e^{-2\eta})^{4/3}}
\equiv \sum^\infty_{k=0} F'_k e^{-2k\eta/3}\,, \nonumber\\
G' &=&- 3 +\frac{1}{1+e^{-2\eta}} +\frac{1}{1+\tilde{H}_0(0)e^{-2\eta}}
\equiv \sum^\infty_{k=0} G'_k e^{-2k\eta/3}
\end{eqnarray}
in which the first few non-vanishing components are
\begin{eqnarray}
&& F'_1=\Lambda_n\ ,\ F'_3=1+\tilde{H}_0(0)\ ,\ F'_4=-\frac{4}{3}\Lambda_n\ ,\ \cdots \nonumber\\
&& G'_0=-1\ ,\ G'_3=-1-\tilde{H}_0(0)\ ,\ \cdots
\end{eqnarray}
With these coefficients we can work out the expansion of $\psi_n$
\begin{eqnarray}
\psi_n(Z) \sim \frac{1}{Z} + \frac{\beta'_1}{Z^{5/3}} + \frac{\beta'_2}{Z^{7/3}}
+ \frac{\beta'_3}{Z^2} + \cdots
\end{eqnarray}
where
\begin{eqnarray}
\beta'_1=-\frac{9}{10}\Lambda_n\ ,\ \beta'_2=\frac{81}{280}\Lambda_n^2\ ,
\ \beta'_3=-\frac{1+\tilde{H}_0(0)}{6}-\frac{27}{560}\Lambda_n^3\ ,\ \cdots
\end{eqnarray}

\begin{figure}[!h]
\includegraphics[width=.4\textwidth]{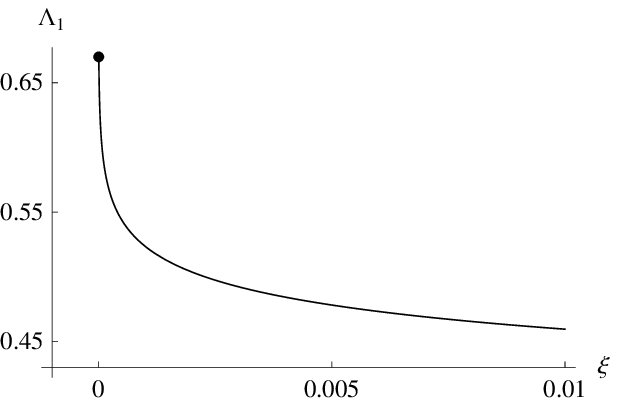}
\includegraphics[width=.4\textwidth]{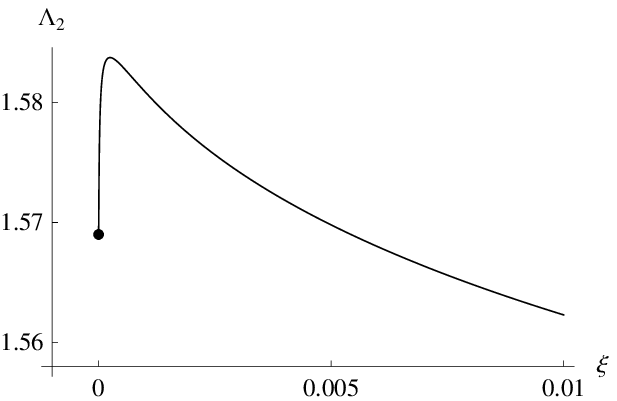}
\caption{The $\xi$-dependence of $\Lambda_1$ and $\Lambda_2$}
\label{fg:LAMBDA}
\end{figure}

\begin{figure}[!h]
\includegraphics[width=.4\textwidth]{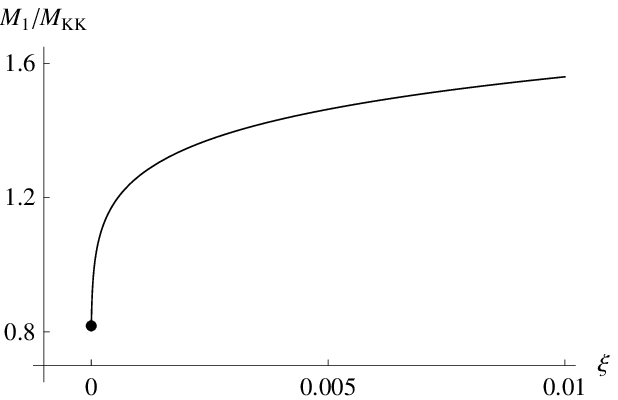}
\includegraphics[width=.4\textwidth]{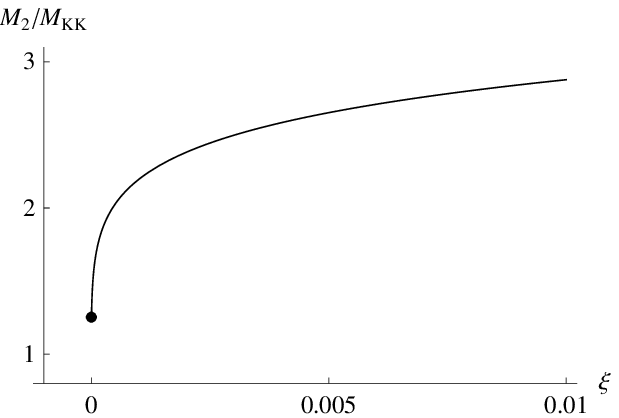}
\caption{The $\xi$-dependence of $M_1$ and $M_2$}
\label{fg:MASS}
\end{figure}

Using shooting method to solve this two-point boundary value problem,
we obtain the evolutions of the first two eigenvalues with respect to $\xi$, which
are shown in Figure~\ref{fg:LAMBDA}, and the corresponding masses are shown in Figure~\ref{fg:MASS}.

 Similar to the scalar meson cases, the contributions from the
eigenvalues are not comparable to the one from the $H_0(U_{KK})$
factor: the trend of the eigenvalues for large $\xi$ is going downward  while
the final masses are going upward.
This result is also different from the D(-1)-D3 case \cite{Brevik:2005fs} in which the
vector mass is independent of $q$.

\begin{figure}[!h]
\includegraphics[width=.4\textwidth]{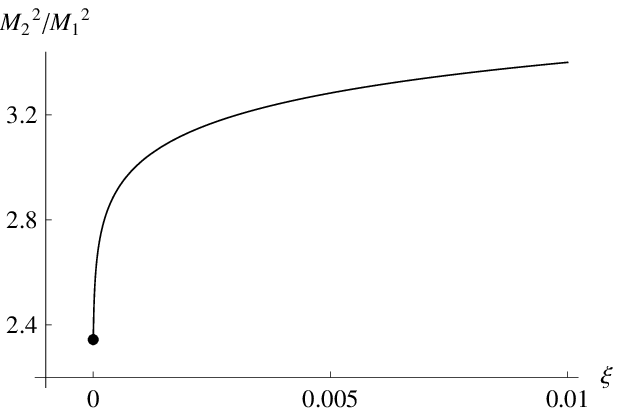}
\includegraphics[width=.4\textwidth]{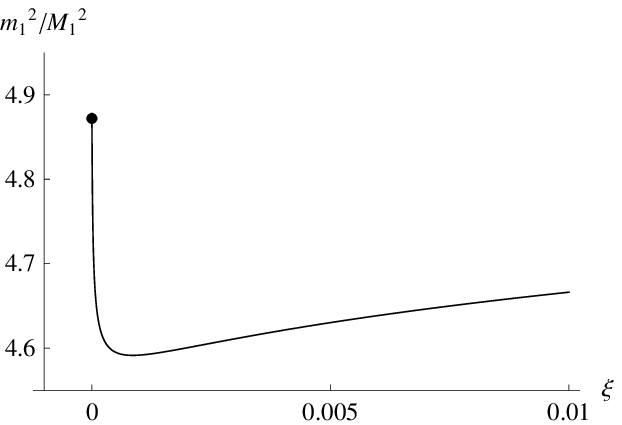}
\caption{The $\xi$-dependence of $M^2_2/M^2_1$ and $m^2_1/M^2_1$}
\label{fg:ratios}
\end{figure}

As Sakai and Sugimoto did in their original paper \cite{Sakai:2004cn}, we could also consider the mass ratios
\begin{eqnarray}
\frac{M^2_2}{M^2_1}=\frac{\Lambda_2}{\Lambda_1} \ ,
\qquad \frac{m^2_1}{M^2_1}=\frac{\lambda_1}{\Lambda_1}\,.
\end{eqnarray}
With the lowest two vector mesons assigned to $\rho(770)$ and
$a_1(1260)$, and the lowest-lying scalar assigned to isospin one $a_0(1450)$, these two
ratios can be estimated to be 2.51 and 3.61 \cite{Sakai:2004cn},
respectively. Our results for these ratios are plotted in Figure~\ref{fg:ratios}.
It is interesting to see that the first estimated ratio can be reached by tuning the $\tilde{\kappa}$-parameter to
some certain value, and the second ratio in our result is closer to the experimental
value with $\tilde \kappa$ turned on. However this should not be
taken seriously, since the experimental value is in the true vacuum in
which the condensate $\langle tr (F_{\mu\nu}\tilde F^{\mu\nu})\rangle$ may be almost zero.

\section {Multiflavor case \label{sect:Nonabelian}}
As in S-S model, we can extend the previous discussion to  the
multi-flavor case, i.e. $N_f>1$. We will see that the mass formulae
for vector mesons are the same as in the one flavor case, and there is no new
information for vector meson mass spectrum. However, we can study the $\tilde
\kappa$ dependence of $f_\pi$ and the couplings
of vectors and Goldstones. Since we closely follow
S-S's original paper in the deduction, we will be brief and refer the readers to
S-S's paper \cite{Sakai:2004cn}.

In multiflavor case, the gauge fluctuations on $N_f$ flavor D8 branes
are nonabelian and the DBI action becomes
\begin{eqnarray}
S_{\rm D8} = -\widehat{T}(2\pi\alpha')^2 \int d^4xdz\
2\,H_0^{1/2}(U_z) {\rm Tr}\left[
\frac{1}{4} \frac{R^3}{U_z} F_{\mu\nu}F^{\mu\nu} +
\frac{9}{8} \frac{U^3_z}{U_{KK}} F_{z\mu}F^{z\mu} \right]
\label {eq:nonabel-DBI}
\end{eqnarray}
where
\begin{eqnarray}
\widehat{T} \equiv	\frac{\tilde{T}}{2} =
\frac{1}{3g_s}\, T_8\, \Omega_4\, U^{1/2}_{KK}\, R^{3/2}
= \frac{M_{KK} N_c \tilde{H}^{1/2}_0(0)}{432\pi^5 \ell_s^6}
\end{eqnarray}
with field strength $F_{MN}=\partial _{M}A_N-\partial_N A_M+[A_M,A_M]$
for $U(N)$ gauge field $A_M$ on the D8 branes. The contractions for
$\mu$ and $\nu$ are done by using $\eta^{\mu\nu}$.

The $U(x^\mu)\equiv\exp\{2i\pi/f_\pi\}$ field in the usual chiral Lagrangian
is realized as 
\begin{equation}
U(x^\mu)=P\exp\Big\{ -\int_{-\infty}^{\infty}dz'A_z(x^\mu,z')\Big\}=\xi_+^{-1}(x^\mu)\xi_{-}(x^\mu) \,
\label{eq:U}
\end{equation}
where $\xi_\pm^{-1}(x^\mu)\equiv
P\exp\{-\int_{0}^{\pm\infty}dz'A_z(x^\mu,z')\}$ is defined for
convenience.

\subsection{Pion Lagrangian}
In $\xi_-(x^\mu)=1$ gauge, one can expand non-Abelian gauge field as
\begin{eqnarray}
A_\mu(x^\mu,z) = U^{-1}(x^\mu) \partial_\mu U(x^\mu) \frac{1+\widehat{\psi}_0(z)}{2}
+ \sum_{n\geq 1} B_\mu^{(n)}(x^\mu) \psi_n(z)
\end{eqnarray}
where
\begin{eqnarray}
\widehat{\psi}_0(z) &=&\frac{ \int_0^ ZdZ \tilde
H_0^{-1/2}(Z)K^{-1}(Z)}{ \int_0^\infty dZ \tilde
H_0^{-1/2}(Z)K^{-1}(Z)}
= \frac{1}{ {\cal F}_0 }\ {\cal F} \left( \arctan\frac{z}{U_{KK}} ,
\sqrt{1- \tilde{H}^{-1}_0(0)} \right)\,.
\end{eqnarray}
with ${\cal F}$ and ${\cal F}_0$ the elliptic integrals defined in
(\ref{eq:elliptic}). Now since we are only interested in the pion field, 
all the excited vector modes $B_\mu^{(n)}$ $(n\ge1)$ can be omitted and
the field strength can be written as 
\begin{eqnarray}
F_{\mu\nu} = \big[U^{-1}\partial_\mu U,U^{-1}\partial_\nu U\big]\
\frac{\widehat{\psi}^2_0 -1}{4}\,,\quad  
F_{z\mu} = U^{-1}\partial_\mu U\ \frac{\partial_z
\widehat{\psi}_0}{2}\,.
\label{eq:pion-field-strength}
\end{eqnarray}

\begin{figure}[!h]
\includegraphics[width=.5\textwidth]{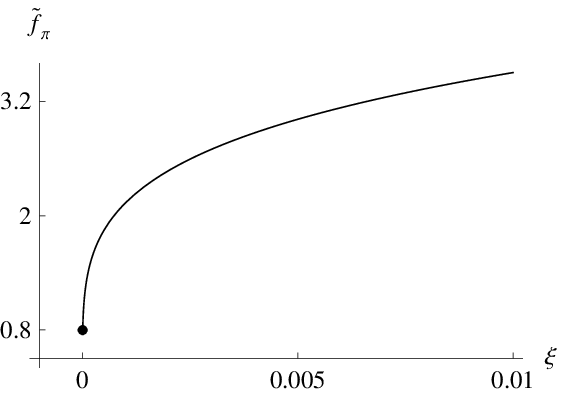}
\includegraphics[width=.5\textwidth]{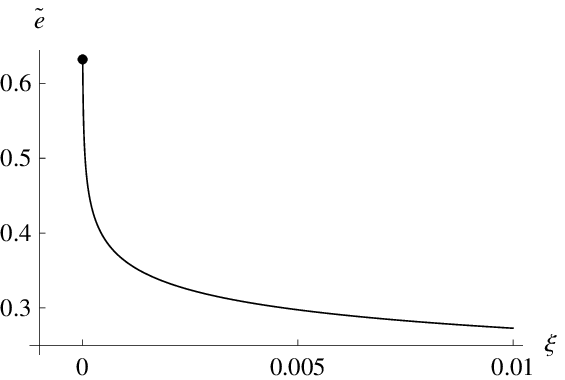}
\caption{The $\xi$-dependence of $\tilde{f}_\pi$ and $\tilde{e}$}
\label{fg:fpi-e}
\end{figure}

Substituting  (\ref{eq:pion-field-strength}) into
(\ref{eq:nonabel-DBI}), we obtain the effective  action for pion
\begin{eqnarray}
S_{\rm D8} = -\widehat{T}(2\pi\alpha')^2 \int d^4x\ {\rm Tr} \Big(
A\,\big(U^{-1}\partial_\mu U\big)^2 +
B\,\big[U^{-1}\partial_\mu U,U^{-1}\partial_\nu U\big]^2 \Big)
\end{eqnarray}
with
\begin{eqnarray}
A &\equiv & \int dz\ \frac{9}{4} \frac{U_z^3}{U_{KK}} H^{1/2}_0(U_z)
\Big( \frac{\partial_z\widehat{\psi}}{2} \Big)^2
= \frac{9U_{KK}\tilde{H}_0^{1/2}(0)}{8{\cal F}_0}\,, \nonumber\\
B &\equiv & \int dz\ \frac{R^3}{2U_z} H_0^{1/2}(U_z)
\Big( \frac{\widehat{\psi}_0^2-1}{4} \Big)^2
=\frac{R^3}{32{\cal F}_0^4}\ b\left(\tilde{H}_0(0)\right)\,.
\end{eqnarray}
Here $b(\tilde{H}_0)$ is an integral constant defined by
\begin{eqnarray}
b(\alpha) \equiv \int dZ\ \frac{(\alpha+Z^2)^{1/2}}{(1+Z^2)^{5/6}}
\left[ {\cal F}^2\Big(\arctan Z,\sqrt{1-\frac{1}{\alpha}}\Big) - {\cal
F}^2_0 \right]^2\,.
\end{eqnarray}
Comparing this result with the Skyrme model \cite{Zahed:1986qz} in which the action is
\begin{eqnarray}
S = \int d^4x\ {\rm Tr}\left( \frac{f_\pi^2}{4}\big(U^{-1}\partial_\mu U\big)^2 +
\frac{1}{32e^2}\big[U^{-1}\partial_\mu U,U^{-1}\partial_\nu U\big]^2
\right)\,,
\end{eqnarray}
we can read off $f^2_\pi$ and the dimensionless $e^2$ 
\begin{eqnarray}
f_\pi^2 &=& 4\widehat{T}(2\pi\alpha')^2 A =
\frac{\tilde{H}_0^2(0)}{108\pi^3{\cal F}_0}\,\lambda N_c M_{KK}^2\,,
 \nonumber\\
e^2 &=& \frac{1}{32\widehat{T}(2\pi\alpha')^2 B} =
\frac{216\pi^3{\cal F}_0^4}{\tilde{H}_0(0)\, b\big(\tilde{H}_0(0)\big)}
\frac{1}{\lambda N_c}\,.
\label{eq:f-pi-e}
\end{eqnarray}
Now the pion decay constant $f_\pi$ and $e$ are both affected
by the glue condensate $\tilde{\kappa}$. 
As before, we use $\xi$ defined in (\ref{eq:xi}) instead of $\tilde \kappa$ and the $\xi$ dependence of $f_\pi$ and $e$ is shown in
Figure (\ref{fg:fpi-e}) where we have defined
\begin{eqnarray}
\tilde{f}_\pi \equiv  \sqrt{\frac{108\pi^3}{\lambda N_c M_{KK}^2}}\,
f_\pi = \frac{\tilde{H}_0(0)}{\sqrt{{\cal F}_0}}\,, \quad
\tilde{e} \equiv  \sqrt{\frac{\lambda N_c}{216\pi^3}} e
=\frac{{\cal F}_0^2}{\sqrt{ \tilde{H}^{1/2}_0(0)\, b\big(\tilde{H}_0(0)\big) }}
\end{eqnarray}
for convenience. We can see that $f_\pi$ goes up while $e$ declines 
 with $\xi$.

\subsection{Vector mesons}
\begin{figure}[!h]
\includegraphics[width=.5\textwidth]{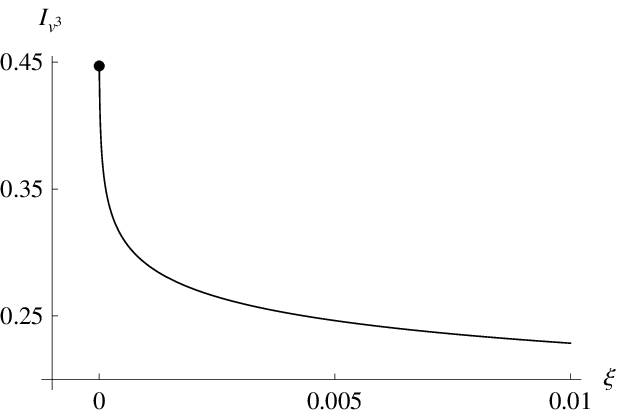}
\includegraphics[width=.5\textwidth]{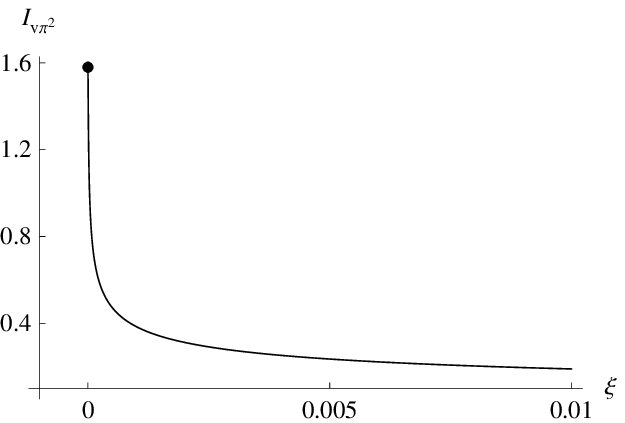}
\caption{The $\xi$-dependence of $I_{v^3}$ and $I_{v\pi^2}$}
\label{fg:couplings}
\end{figure}

Next we consider the first excited vector mode $B_\mu^{(1)}$ which is identified as the $\rho$ meson. 
In  the $\xi_+^{-1}(x^\mu)=\xi_-(x^\mu)=\exp(i\pi(x^\mu)/f_\pi)$ gauge,
$A_\mu$ can be expanded as
\begin{eqnarray}
A_\mu(x^\mu,z) = \frac{i}{f_\pi} \partial_\mu \pi(x^\mu) \widehat{\psi}_0(z)
+ \frac{1}{2f_\pi^2} \big[ \pi(x^\mu), \partial_\mu \pi(x^\mu) \big]
+ v_\mu(x^\mu) \psi_1(z)
\end{eqnarray}
where $v_\mu=B_\mu^{(1)}$. Thus the field strength is 
\begin{eqnarray}
F_{\mu\nu} &=& \frac{i}{f_\pi}\left(
\big[ \partial_\mu\pi,v_\nu \big] + \big[ v_\mu,\partial_\nu\pi \big]
\right) \psi_1 \widehat{\psi}_0 + \frac{1}{f_\pi^2}
\big[ \partial_\mu\pi,\partial_\nu\pi \big] \big( 1-\widehat{\psi}_0^2 \big) \nonumber\\
&& + (\partial_\mu v_\nu - \partial_\nu v_\mu) \psi_1
+ \big[ v_\mu , v_\nu \big] \psi_1^2 + {\cal O}\big( (\pi,v_\mu)^3
\big)\,, 
\nonumber\\
F_{z\mu} &=& \frac{i}{f_\pi} \partial_\mu\pi\, \partial_z\widehat{\psi}_0
+ v_\mu\, \partial_z \psi_1\,.
\end{eqnarray}
The effective action involving $\pi$ and $v_\mu$ up to ${\cal O}\big(
(\pi,v_\mu)^3\big)$ can be obtained  
\begin{eqnarray}
S_{\rm D8} &=& \int d^4x\ \left\{
- a_{\pi^2} {\rm Tr}\big( \partial_\mu \pi \partial^\mu \pi \big)
 + a_{v^2} \Big( {\rm Tr}\big( \partial_\mu v_\nu -
\partial_\nu v_\mu \big)^2 + m_v^2\,{\rm Tr}\,v_\mu^2 \Big)\right. \nonumber\\
&& +\ a_{v^3} {\rm Tr} \Big( \big[ v^\mu , v^\nu \big]
\big( \partial_\mu v_\nu - \partial_\nu v_\mu \big) \Big)
 + a_{v\pi^2} {\rm Tr} \Big( \big[ \partial^\mu\pi , \partial^\nu\pi \big]
\big( \partial_\mu v_\nu - \partial_\nu v_\mu \big) \Big) \nonumber\\
&& \left. +\ {\cal O}\big( (\pi,v_\mu)^4 \big) \right\}\,.
\end{eqnarray}
Then we determine all the coefficients one by one.  The coefficient
before the kinetic term of pion is 
\begin{eqnarray}
a_{\pi^2} = \frac{2\widehat{T}(2\pi\alpha')^2}{f_\pi^2}
\int dz\ \frac{9}{8} \frac{U_z^3}{U_{KK}} H_0^{1/2}(U_z)
\big( \partial_z\widehat{\psi}_0 \big)^2 =1
\end{eqnarray}
due to the definition of $f_\pi$ in (\ref{eq:f-pi-e}). Next we redefine
\begin{eqnarray}
\Psi_1(Z) \equiv \sqrt{\widehat{T}(2\pi\alpha')^2 R^3}\, \psi_1(U_{KK}Z)
\end{eqnarray}
so that it is properly normalized and the coefficient before the
vector kinetic term is 
\begin{eqnarray}
a_{v^2} &=& \widehat{T}(2\pi\alpha')^2 \int dz\ \frac{R^3}{U_z}
H_0^{1/2}(U_z) \psi_1^2(z) \nonumber\\
&\equiv & \int dZ\ K^{-1/3}(Z) \tilde{H}^{1/2}_0(Z) \Psi_1^2(Z) =1
\end{eqnarray}
by the orthogonal  condition (\ref{eq:ortho0}). This leads to
\begin{eqnarray}
m_v^2 = a_{v^2}\,m_v^2 &=& \widehat{T}(2\pi\alpha')^2
\int dz\ \frac{9}{4} \frac{U_z^2}{U_{KK}} H_0^{1/2}(U_z)
\Big( \frac{d\psi_1(z)}{dz} \Big)^2 \nonumber\\
&=& \Lambda_1 M_{KK}^2 \tilde{H}_0(0) \equiv m_1^2
\end{eqnarray}
which is in agreement with equation (\ref{eq:phi}) except for a
redefinition of $\widehat{T}$. So the $\tilde \kappa$ dependence of
the vector mass is the same as in the one flavor case.
The three-point self-coupling for the vector field is
\begin{eqnarray}
a_{v^3} &=& \widehat{T}(2\pi\alpha')^2 \int dz\ \frac{R^3}{U_z}
H_0^{1/2}(U_z) \psi_1^3(z) \nonumber\\
&=& \frac{(6\pi)^{3/2}}{\sqrt{\lambda N_c}}\,
I_{v^3} \big( \tilde{H}_0(0) \big)\,.
\end{eqnarray}
Similarly, the vector-Goldstone-Goldstone(VGG) three-point coupling is
\begin{eqnarray}
a_{v\pi^2} &=& \frac{\widehat{T}(2\pi\alpha')^2}{f_\pi^2}
\int dz\ \frac{R^3}{U_z} H_0^{1/2}(U_z)
\psi_1 \big( 1-\widehat{\psi}_0^2 \big) \nonumber\\
&=& \frac{\pi(3\pi)^{3/2}}{M_{KK}^2 \sqrt{2\lambda N_c}}\,
I_{v\pi^2} \big( \tilde{H}_0(0) \big)\,.
\end{eqnarray}
Here we have defined
\begin{eqnarray}
I_{v^3}\big(\tilde{H}_0(0)\big) &=& \frac{1}{\tilde{H}^{1/4}_0(0)} \int dZ\
\frac{\big(\tilde{H}_0(0)+Z^2\big)^{1/2}}{(1+Z^2)^{5/6}}
\Psi_1^3(Z)\,,
 \nonumber\\
I_{v\pi^2}\big(\tilde{H}_0(0)\big) &=& \frac{2{\cal F}_0}{\pi \tilde{H}^{7/4}_0(0)}
\int dZ\ \left[ 1- \frac{1}{{\cal F}_0^2}
{\cal F}^2\Big(\arctan Z,\sqrt{1-\tilde{H}_0^{-1}(0)}\,\Big) \right] \nonumber\\
&& \qquad\qquad\qquad\qquad
\times\ \frac{\big(\tilde{H}_0(0)+Z^2\big)^{1/2}}{(1+Z^2)^{5/6}}
\Psi_1(Z)\,.
\label{eq:I}
\end{eqnarray}
We can see that the couplings depend on $\tilde \kappa$ both
explicitly in 
$\tilde{H}_0(0)$ in the integrands and the coefficients before the
integrals, and implicitly in
eigenfunction $\psi_1$ through the appearance of $\tilde H_0(Z)$ in the eigenvalue equation.
 The
 dependence is illustrated in Figure~\ref{fg:couplings}. So
both the three point self-interaction of the vector meson and the
VGG coupling are becoming weaker when $\tilde \kappa$ is turned on. 

\section{Conclusion and discussion\label{sect:conclude}}

In this paper, we have studied the S-S model in the D0-D4 background. The
corresponding gauge field theory has a nonzero condensate $\langle
tr (F_{\mu\nu}\tilde F^{\mu\nu})\rangle$. The effects of this quantity
on the meson spectra, pion decay constant and the lowest-lying 
three-vector and  vector-Goldstone-Goldstone couplings are 
studied. The dependence of these quantities on $\tilde \kappa$ comes
in two parts: one is an explicit dependence in the $H_0(U_{KK})$
factor in the formulae and the other is implicit in the eigenvalues
and eigenfunctions (see (\ref{eq:Mass-Scalar}), (\ref{eq:eigeneq}) and
(\ref{eq:I})). The $\tilde \kappa$ dependence of the mass spectra are
different from the D(-1)-D3 case in \cite{Brevik:2005fs}.
On the gravity side, this
$\tilde\kappa$ dependence  comes from  the backreaction of the $D0$
charges to the metric and takes effect by coupling the metric to 
the flavor branes in the DBI action. On the  field theory side, the glue
condensate comes into play through its backreaction on the glue
fluctuations which couple to the flavors through glue-quark
couplings. It must be a strong coupling
nonperturbative phenomenon to have sizable effects on the mass spectra
and the couplings.  So these  two pictures seem to be consistent.
However, the $\tilde \kappa$ dependence in the metric always appears in
$\tilde \kappa^2$, the squared form. The Chern-Simons terms with $C_1$ form field for D8 is zero since it
involves fluctuations in the $S_4$ directions.   So there is
no explicit $P/CP$ breaking terms in the 
effective Lagrangian from DBI action --- no $P/CP$ violating mixings
and interactions, which seems to be strange since $P/CP$ is
 broken due to nonzero condensate $\tilde \kappa$. One tends to give a
handwaving argument as follows: Since the condensate is in pure glue
sector, there could be $P/CP$ violating mixings of glueballs
when $\tilde \kappa$ is nonzero. The $P/CP$
violating mixings of mesons in this model only happen through
intermediate glueball mixing, and due to the OZI rule, this process may be
suppressed in the large $N_c$.

In this paper we have not studied the Chern-Simons term containing
$f_4$. This term can produce more interaction terms
\cite{Sakai:2005yt} and is also related to the baryons in this model
\cite{Hata:2007mb,Hashimoto:2008zw}.  It is easy to extend the
discussions to this term in the D0-D4 background, which allows one to
learn more interactions and the baryon properties with regard to the
condensate $\tilde \kappa$. However, to introduce deconfinement
temperature into this model is a little difficult since this needs a
background with a horizon in the four-spacetime also with the form
field $C_\tau d\tau$, which may not be easy to find. 

The string theory background used in this paper corresponds to a gauge
theory with a {\em real} Euclidean condensate $\langle tr (F_{\mu\nu}\tilde
F^{\mu\nu})\rangle$ as in \cite {Liu:1999fc} in which the gauge theory
background was claimed to be selfdual. Since our background here is
not supersymmetric, the selfdual property is not clear at present.
Considering the quantum effect there could be some modification to
$\tilde \kappa$. The situation that a nonextremal gravity background
can lead to a non-selfdual
field strength has been studied in \cite{Bergshoeff:2005zf} in the context
of the localized instanton case. We leave this direction for future research. The
absolute value of the quantities studied in this paper may not be of much
significance. But the tendency of these quantities as $\tilde\kappa$ is
turned on may capture the qualitative effect of the real Euclidean
condensate in this model.  However, a real Euclidean condensate may not be realistic
in the real world which is Minkowski. The string theory background for
gauge theory with real Minkowski condensates can also be found.
However there could arise some other problems. We are still working on
this possibility. The preliminary result is that, the real Minkowski
condensate may have opposite effects on the quantities studied in this
paper to the Euclidean one.  This needs to be confirmed in future
work.

\section*{Acknowledgements}
This work is supported by the NSF of China under grant No.11105138 and
11235010 
and is also supported by the Fundamental Research Funds for the
Central Universities under grant No.WK2030040020. We also thank S.~-J.~Sin
for helpful discussion.


\begin{thebibliography}{99}
\bibitem{Simonov:1997js}
  Yu.~A.~Simonov,
  Phys.\ Usp.\  {\bf 39} (1996) 313
   [Usp.\ Fiz.\ Nauk {\bf 166} (1996) 337]
  [hep-ph/9709344].

\bibitem{Leutwyler:1980ev}
  H.~Leutwyler,
  Phys.\ Lett.\ B {\bf 96} (1980) 154;
  Nucl.\ Phys.\ B {\bf 179} (1981) 129.


\bibitem{Minkowski:1981ma} 
  P.~Minkowski, 
  Nucl.\ Phys.\ B {\bf 177} (1981) 203. 


\bibitem{Flory:1983dx}
  C.~A.~Flory,
  Phys.\ Rev.\ D {\bf 28} (1983) 1425.

\bibitem{vanBaal:1984ar}
  P.~van Baal,
  Commun.\ Math.\ Phys.\  {\bf 94} (1984) 397.

\bibitem{Efimov:1998hi}
  G.~V.~Efimov, A.~C.~Kalloniatis and S.~N.~Nedelko,
  Phys.\ Rev.\ D {\bf 59} (1999) 014026
  [hep-th/9806165].


\bibitem{Kharzeev:1998kz}
  D.~Kharzeev, R.~D.~Pisarski and M.~H.~G.~Tytgat,
  Phys.\ Rev.\ Lett.\  {\bf 81} (1998) 512
  [hep-ph/9804221];
  hep-ph/9808366;
  [hep-ph/0012012].


\bibitem{Buckley:1999mv}
  K.~Buckley, T.~Fugleberg and A.~Zhitnitsky,
  Phys.\ Rev.\ Lett.\  {\bf 84} (2000) 4814
  [hep-ph/9910229].

\bibitem{Kharzeev:2004ey}
  D.~Kharzeev,
  Phys.\ Lett.\ B {\bf 633} (2006) 260
  [hep-ph/0406125].


\bibitem{Liu:1999fc}
  H.~Liu and A.~A.~Tseytlin,
  Nucl.\ Phys.\ B {\bf 553} (1999) 231
  [hep-th/9903091].

\bibitem{Kehagias:1999iy}
  A.~Kehagias and K.~Sfetsos,
  Phys.\ Lett.\ B {\bf 456} (1999) 22
  [hep-th/9903109].

\bibitem{Ghoroku:2005tf}
  K.~Ghoroku, T.~Sakaguchi, N.~Uekusa and M.~Yahiro,
  Phys.\ Rev.\ D {\bf 71} (2005) 106002
  [hep-th/0502088].

\bibitem{Brevik:2005fs}
  I.~H.~Brevik, K.~Ghoroku, A.~Nakamura and ,
  Int.\ J.\ Mod.\ Phys.\ D {\bf 15} (2006) 57
  [hep-th/0505057].

\bibitem{Karch:2002sh}
  A.~Karch and E.~Katz,
  JHEP {\bf 0206} (2002) 043
  [hep-th/0205236].

\bibitem{Ghoroku:2004sp}
 K.~Ghoroku and M.~Yahiro,
  Phys.\ Lett.\ B {\bf 604} (2004) 235
  [hep-th/0408040].
\bibitem{Ghoroku:2006af}
  K.~Ghoroku, M.~Ishihara and A.~Nakamura,
  Phys.\ Rev.\ D {\bf 74} (2006) 124020
  [hep-th/0609152].

\bibitem{Erdmenger:2007vj}
  J.~Erdmenger, K.~Ghoroku and I.~Kirsch,
  JHEP {\bf 0709} (2007) 111
  [arXiv:0706.3978 [hep-th]].
\bibitem{Erdmenger:2011sz}
  J.~Erdmenger, A.~Gorsky, P.~N.~Kopnin, A.~Krikun and A.~V.~Zayakin,
  JHEP {\bf 1103} (2011) 044
  [arXiv:1101.1586 [hep-th]].

\bibitem{Ghoroku:2008tg}
  K.~Ghoroku and M.~Ishihara,
  Phys.\ Rev.\ D {\bf 77} (2008) 086003
  [arXiv:0801.4216 [hep-th]].

\bibitem{Ghoroku:2008na}
  K.~Ghoroku, M.~Ishihara, A.~Nakamura and F.~Toyoda,
  Phys.\ Rev.\ D {\bf 79} (2009) 066009
  [arXiv:0806.0195 [hep-th]].

\bibitem{Sin:2009yu}
  S.~-J.~Sin, S.~Yang and Y.~Zhou,
  JHEP {\bf 0911} (2009) 001
  [arXiv:0907.1732 [hep-th]].

\bibitem{Gwak:2012ht}
  B.~Gwak, M.~Kim, B.~-H.~Lee, Y.~Seo and S.~-J.~Sin,
  Phys.\ Rev.\ D {\bf 86} (2012) 026010
  [arXiv:1203.4883 [hep-th]].

\bibitem{Sin:2009dk}
  S.~-J.~Sin and Y.~Zhou,
  JHEP {\bf 0905} (2009) 044
  [arXiv:0904.4249 [hep-th]].

\bibitem{Witten:1998zw}
  E.~Witten,
  Adv.\ Theor.\ Math.\ Phys.\  {\bf 2} (1998) 505
  [hep-th/9803131].

\bibitem{Kruczenski:2003uq}
  M.~Kruczenski, D.~Mateos, R.~C.~Myers and D.~J.~Winters,
  JHEP {\bf 0405} (2004) 041
  [hep-th/0311270].

\bibitem{Sakai:2004cn}
  T.~Sakai and S.~Sugimoto,
  Prog.\ Theor.\ Phys.\  {\bf 113} (2005) 843
  [hep-th/0412141].

\bibitem{Sakai:2005yt}
  T.~Sakai and S.~Sugimoto,
  Prog.\ Theor.\ Phys.\  {\bf 114} (2005) 1083
  [hep-th/0507073].

\bibitem{Hata:2007mb}
  H.~Hata, T.~Sakai, S.~Sugimoto and S.~Yamato,
  Prog.\ Theor.\ Phys.\  {\bf 117} (2007) 1157
  [hep-th/0701280 [HEP-TH]].

\bibitem{Hashimoto:2008zw}
  K.~Hashimoto, T.~Sakai and S.~Sugimoto,
  Prog.\ Theor.\ Phys.\  {\bf 120} (2008) 1093
  [arXiv:0806.3122 [hep-th]].

\bibitem{Hashimoto:2009ys}
  K.~Hashimoto, T.~Sakai and S.~Sugimoto,
  Prog.\ Theor.\ Phys.\  {\bf 122} (2009) 427
  [arXiv:0901.4449 [hep-th]].

\bibitem{Kaplunovsky:2010eh}
  V.~Kaplunovsky and J.~Sonnenschein,
  JHEP {\bf 1105} (2011) 058
  [arXiv:1003.2621 [hep-th]].

\bibitem{Barbon:1999zp}
  J.~L.~F.~Barb\'on and A.~Pasquinucci,
  Phys.\ Lett.\ B {\bf 458} (1999) 288
  [hep-th/9904190].
\bibitem{Suzuki:2000sv}
  K.~Suzuki,
  Phys.\ Rev.\ D {\bf 63} (2001) 084011
  [hep-th/0001057].

\bibitem{Zahed:1986qz}
  I.~Zahed, G.~E.~Brown and ,
  Phys.\ Rept.\  {\bf 142} (1986) 1.


\bibitem{Bergshoeff:2005zf}
  E.~Bergshoeff, A.~Collinucci, A.~Ploegh, S.~Vandoren and T.~Van Riet,
  JHEP {\bf 0601} (2006) 061
  [hep-th/0510048].
\end{thebibliography}
\end{document}